\documentclass[aps,prl,twocolumn,superscriptaddress,showpacs]{revtex4}

\usepackage{txfonts}
\usepackage{amssymb}
\usepackage{graphicx}

\begin{document}

\title{Constrain on possible pairing symmetry in a two-orbital model of
FeAs-based superconductors}
\author{Wen-Long You}
\affiliation{Department of Physics and ITP, The Chinese University of Hong
Kong, Hong Kong, China}%
\affiliation{School of Physics, Peking University, Beijing, 100871, China}
\author{Shi-Jian Gu}
\affiliation{Department of Physics and ITP, The Chinese University of Hong
Kong, Hong Kong, China}
\author{Guang-Shan Tian}
\affiliation{Department of Physics and ITP, The Chinese University of Hong
Kong, Hong Kong, China}%
\affiliation{School of Physics, Peking University, Beijing, 100871, China}
\author{Hai-Qing Lin}
\affiliation{Department of Physics and ITP, The Chinese University of Hong
Kong, Hong Kong, China}

\begin{abstract}
In this work, we establish a few exact identities through commutation of
intra-orbital and inter-orbital on-site pairings with a two-orbital model
describing newly discovered FeAs-based superconductors. Applying the conclusion
drawn from rigorous relation and physical interpretation, we give constraints
on the possible symmetries of the superconducting pairing of the model. Hence
the favorable pairings in newly discovered high-temperature oxypnictide
superconductors are proposed.
\end{abstract}


\date{\today}
\maketitle




\emph{Introduction}---The newly discovered family of FeAs-based ReO$_{1-x}$F$%
_{x}$FeAs (Re=La, Nd, Ce, Pr, etc.) high-temperature superconductors (SC)%
\cite{Kamihara-JACS-2008,KakahashiNature} spark great interests both
experimentally \cite%
{ZXZhao,NLWang,ZhuanXu,HHWen0,DFFang08033603,DLFeng08034328,Cruz-arXiv08040795,
Ahilan-arXiv08044026,Nakai-arXiv08044765,Matano-arXiv08060249,HHWen}
and theoretically \cite%
{DJSingh,Halule-PRL-2008,arXiv08050736,arXiv08050923,arXiv08060259,
arXiv08060712,Mazin-arXiv08032740,Kuroki-arXiv-2008,Korshunov-arXiv-2008,
Dai-arXiv08033982,Cvetkovic-arXiv-2008,Wen-arXiv08041739,Xu-EPL-2008,
ZDWang-arXiv-2008,TLi-arXiv-2008,QSi-arXiv-2008,Han-EPL-2008,ZYWeng-arXiv-2008,
Qi-arXiv-2008,SYang}.%
So far the transition temperature $T_{c}$ has gone up as high as above $50$ K
\cite{ZXZhao,NLWang,ZhuanXu}. The new family material provides another platform
to explore high-Tc superconductivity besides cuprate superconductors. Both
FeAs-based materials and cuprate superconductors are transition-metal compounds
on a two-dimensional (2D) square lattice, and their parent compounds are
magnetically ordered. However, there are several significant differences in
electronic properties. Firstly, the undoped oxypnictides are bad metals
\cite{NLWang,DJSingh}, while Cuprates are Mott insulators. Secondly, neutron
scattering experiments have shown that the magnetic structure in undoped
oxypnictides is not the same as a simple antiferromagnetic order in cuprates
but instead a collinear spin-density wave along the ($\pi$, 0) direction
\cite{NLWang,Cruz-arXiv08040795}. Thirdly, probably the most important,
multi-orbital nature of the oxypnictides has been emphasized, in contrast to
the single-band cuprates. From the band structure point of view, it seems
likely that all $3d$ orbitals of the Fe atoms are involved in the low energy
electronic properties\cite{DJSingh,Halule-PRL-2008}.

Meanwhile, the pairing symmetry remains controversial. Based on fermionic
nature of the gap function, the possible superconducting order parameters can
be classified according to the group theory
\cite{arXiv08050736,arXiv08050923,arXiv08060259,arXiv08060712}. There are
surveys that support either $s$- or extended $s$-wave\cite%
{Dai-arXiv08033982,Cvetkovic-arXiv-2008,Mazin-arXiv08032740,Kuroki-arXiv-2008,
Korshunov-arXiv-2008},%
or $p$-wave \cite{Wen-arXiv08041739,Xu-EPL-2008}, or $d$-wave \cite%
{ZDWang-arXiv-2008,TLi-arXiv-2008,QSi-arXiv-2008,Han-EPL-2008,ZYWeng-arXiv-2008,
Qi-arXiv-2008},%
even mixture of $s_{xy}$ and $d_{x^2-y^2}$ \cite{Kangjun-arXiv08052958,SYang}.
Nevertheless, none of them is confirmative. Recently a nodal superconductivity
in the electron-doped oxypnictides with multiple gaps structure was suggested
by specific heat \cite{HHWen0} as well as the nuclear-magnetic-resonance(NMR)
experiments \cite{Ahilan-arXiv08044026,Nakai-arXiv08044765,
Matano-arXiv08060249} and point-contact spectroscopy \cite{HHWen}. Therefore,
the investigation of possible coexistence of various superconducting orders is
highly desired.

\begin{figure}[h]
\includegraphics[bb=-180 275 725 560, width=8.5cm, clip]{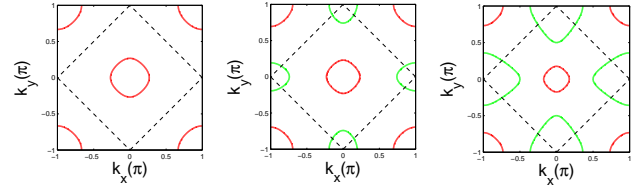}
\caption{ (color online) Top: The Fermi surface of the two-orbital model on
the large 1Fe/cell BZ at $\protect\mu$=1.00, $\protect\mu$=1.45, $\protect%
\mu $=2.00 from left to right. The dashed square indicates the BZ of
2Fe/cell. The parameters are consistent with those in Ref.\cite
{Raghu-PRB-2008}.} \label{fig:FermiSurface}
\end{figure}

\emph{The Hamiltonian}---The Fe atoms in a Fe-As plane form a 2D square
lattice. Due to the buckling of the As atoms, the real unit cell contains two
Fe atoms. As shown from crystal field splitting and simple valence counting as
well as more reliable local-density-approximation(LDA) calculations
\cite{DJSingh}, it is reasonable to assume that Fe $3d_{xz}$ and $3d_{yz}$
orbitals play an important role in the low energy physics of this material,
which are locally degenerate due to the tetragonal symmetry. LDA results also
show the presence of small Fermi surfaces (FS) \cite{Dai-arXiv08033982}. In the
unfolded Brillouin zone consisting of one Fe per unit cell, electron and hole
pockets exist around $M$ and $\Gamma $ points, respectively. The system has
multi-band with a hole-like Fermi surface around the $\Gamma $ point and an
electron-like around the $M$ point of the Brillouin zone. Upon doping, the
hole-like FS shrink rapidly, while the electron FS expand their areas, as shown
in Fig. \ref{fig:FermiSurface}. The dominant contribution is clearly around
$\Gamma $, $M$. Thereby, we consider a 2D square lattice with \textquotedblleft
$d_{xz}$, $d_{yz}$" orbitals per site as a starting model to describe
FeAs-based superconductors \cite{Raghu-PRB-2008} for the oxypnictide compounds,
\begin{eqnarray}
H &=&H_{0}+H_{I}  \label{MainHamiltonian} \\
H_{0} &=&-\sum_{ij\sigma }t_{ij,\sigma }^{cc}(c_{i,\sigma }^{\dagger
}c_{j,\sigma }+\textrm{H.c.})-\sum_{ij\sigma }t_{ij,\sigma
}^{dd}(d_{i,\sigma
}^{\dagger }d_{j,\sigma }+\textrm{H.c.})  \nonumber \\
&&-\sum_{(ij)\sigma }t_{ij,\sigma }^{cd}(c_{i,\sigma }^{\dagger
}d_{j,\sigma }+H.c.)-\mu \sum_{i\sigma }(n_{i,c,\sigma }+n_{i,d,\sigma }) ~, \\
H_{I} &=&\sum_{i}\big(U_{c}n_{i,c,\uparrow }n_{i,c,\downarrow
}+U_{d}n_{i,d,\uparrow }n_{i,d,\downarrow }  \nonumber \\
&&+U_{cd}n_{i,c}n_{i,d}-J_{H}S_{i,c}\cdot S_{i,d}\big) ~.
\end{eqnarray}%
Here $c$ $(d)$ labels $d_{xz}$ $(d_{yz})$ orbital, $c_{i,\sigma}^{\dagger}
(d_{i,\sigma }^{\dagger })$ is creation operator for electrons of spin $%
\sigma $ and orbital $d_{xz}$ $(d_{yz})$ at site $i$, $n_{i,c,\sigma
}=c_{i,\sigma}^{\dagger}c_{i,\sigma }$, $n_{i,d,\sigma
}=d_{i,\sigma}^{\dagger}d_{i,\sigma }$, $\tau _{i}$ is the pauli matrices for
two orbitals at site $i$,  the hoping integrals are
$t_{ij,\sigma }^{cc}=t_{1}\delta (i-j+\hat{x}%
)+t_{2}\delta (i-j+\hat{y})+t_{3}\delta (i-j+\hat{x}+\hat{y})+t_{3}\delta
(i-j-\hat{x}+\hat{y})$, $t_{ij,\sigma }^{dd}=t_{2}\delta (i-j+\hat{x}%
)+t_{1}\delta (i-j+\hat{y})+t_{3}\delta (i-j+\hat{x}+\hat{y})+t_{3}\delta
(i-j-\hat{x}+\hat{y})$ and $t_{ij,\sigma }^{cd}=t_{4}\delta (i-j+\hat{x}-%
\hat{y})+t_{4}\delta (i-j-\hat{x}+\hat{y})-t_{4}\delta (i-j+\hat{x}+\hat{y}%
)-t_{4}\delta (i-j-\hat{x}-\hat{y})$ (as illustrated in Fig. 1 of Ref. \cite%
{Raghu-PRB-2008}), $U_{c},U_{d}$ are intraband Coulomb repulsion with the
relation $U_{c}=U_{d}\equiv U$, $U_{cd}$ interband Coulomb repulsion, and $%
J_{H}$ the Hund's rule coupling. For later convenience, we set
$t_{\perp}=(t_{1}+t_{2})/2$, $t_{\parallel}=(t_{1}-t_{2})/2$. The space group
of Fe atoms is $P4/nmm$, and it is characterized by point group $D_{4}$ and
lattice translation group $T$. Therefore, the basis matrix functions belong to
different irreducible representations of point group $D_{4}$, which has five
irreducible representations, including four one-dimensional representations
($A_{1}$, $A_{2}$, $B_{1}$ and $B_{2}$) and one two-dimensional representation
($E$). Typical bases in each representation are listed in Table \ref{IRofD4h}.
The orbital part of the pairing matrix $\Omega $ is spanned in the vector space
of ($d_{xz}$, $d_{yz}$), which is an irreducible representation $E_{g}$ of the
point group $D_{4}$, and $\tau _{0,1,2,3}$ in Table \ref{IRofD4h} are
transformed as $A_{1}$, $B_{2}$, $A_{2}$, $B_{1}$ respectively
\cite{arXiv08050736,arXiv08050923,arXiv08060712}.

\begin{table}[t]
\caption{One dimensional irreducible representations of $D_{4}$ group in
spatial and orbital space}
\label{IRofD4h}%
\begin{ruledtabular}
\begin{tabular}{ c |  c |  c }
IR & Spatial Basis Functions & Bases in Orbital Space $\Omega$
\\ \hline
$A_{1}$ &$\cos k_x\cos k_y$ , $\cos k_x+\cos k_y$ & $\tau_0$ \\
$B_{2}$ &$\sin k_x\sin k_y$ & $\tau_1$ \\
$A_{2}$ &$\sin k_x\sin k_y (\cos k_x-\cos k_y)$ & $\tau_2$ \\
$B_{1}$ &$\cos k_x-\cos k_y$ & $\tau_3$
\end{tabular}
\end{ruledtabular}
\end{table}

In the followings, we will establish sum rules for various pairings by
exploiting the commutation relations between Hamiltonian and on site pairing
operators. Suppose $A_{i}$, $B_{i}$, $C_{i}$, and $D_{i}$ be some localized
operators defined on lattice $\Lambda $, and $\Psi _{0}$ be an absolute ground
state of $H_{\Lambda }$. If they satisfy the following commutation relation
$[H_{\Lambda },A_{i}]=\alpha B_{i}+\beta C_{i}+\gamma D_{i}
\label{CommunicationEquality100}$, where $\alpha \neq 0$, $\beta \neq 0$,%
and $\gamma \neq 0$ are some constants. Since $\langle \Psi _{0}|[H_{\Lambda
},A_{i}]|\Psi _{0}\rangle =(E_{0}-E_{0})\langle \Psi _{0}|A_{i}|\Psi
_{0}\rangle =0$ so $\alpha \langle \Psi _{0}|B_{i}|\Psi _{0}\rangle +\beta
\langle \Psi _{0}|C_{i}|\Psi _{0}\rangle +\gamma \langle \Psi _{0}|D_{i}|\Psi
_{0}\rangle =0$, hence the orders of $B_{i}, C_{i}$, and $D_{i}$ should be
either absent together or at least two of them exist simultaneously in the
ground state \cite{SCZhang-PRB-1990}. The rigorous proof will be stated later.

\begin{figure}[t]
\includegraphics[width=0.5\textwidth]{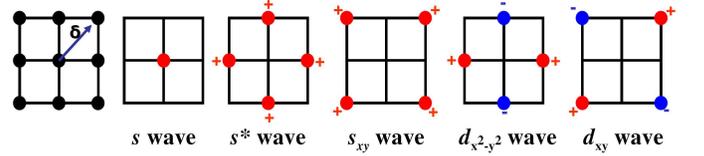}
\caption{ (color online) The pair coordinates $\vec{\protect\delta}$ used to
defining $\Delta_{\vec{\protect\delta}}$, ranges over 5 positions including
(0,0). }
\label{fig:Pairconf}
\end{figure}

\emph{Intra-orbital pairing}---We consider the intra-orbital pairing. Define on
site pairing $\Delta _{r}^{cc}=c_{r,\uparrow }c_{r,\downarrow }$, $\Delta
_{r}^{dd}=d_{r,\uparrow }d_{r,\downarrow }$, nearest-neighbor (NN) or next
nearest-neighbor (NNN) spin singlet intra-orbital pairing
$\Delta_{r+\vec{\delta}}^{cc}=c_{r,\uparrow }c_{r+ \vec{\delta},\downarrow
}-c_{r,\downarrow }c_{r+\vec{\delta},\uparrow }$, $\Delta
_{r+\vec{\delta}}^{dd}=d_{r,\uparrow }d_{r+\vec{\delta},\downarrow
}-d_{r,\downarrow }d_{r+\vec{\delta},\uparrow }$,%
and NN or NNN spin singlet inter-orbital pairing%
$\Delta _{r+\vec{\delta} }^{cd}=c_{r,\uparrow }d_{r+\vec{\delta},\downarrow
}-c_{r,\downarrow }d_{r+\vec{\delta},\uparrow }$, $\Delta
_{r+\vec{\delta}}^{dc}=d_{r,\uparrow }c_{r+\vec{\delta},\downarrow
}-d_{r,\downarrow }c_{r+\vec{\delta},\uparrow }$.%
By calculating the commutation relation with the Hamiltonian, which can be
expressed as linear combination of $s$ and $d$ wave symmetry involving NN and
NNN electrons, such as %
on site $s$ wave pairing operator $\Delta _{s}=\sum_r \Delta_r /N$, %
extended $s^{\ast}$ wave pairing operator $\Delta _{s\ast }=\sum_{r}(\Delta
_{r-\hat{x} }+\Delta _{r+\hat{x}}+\Delta _{r-\hat{y}}+\Delta _{r+\hat{y}})/N$, %
${s_{xy}}$ wave pairing operator $\Delta_{s_{xy}} = \sum_{r}
(\Delta_{r+\hat{x}+\hat{y}}+\Delta_{r-\hat{x}-\hat{y}}+
\Delta_{r+\hat{x}-\hat{y}}+\Delta_{r-\hat{x}+\hat{y}})/N$, %
$d_{x^{2}-y^{2}}$ wave pairing operator $\Delta _{d_{x^{2}-y^{2}}} =
\sum_{r}(\Delta_{r-\hat{x}}+\Delta _{r+\hat{x}}-\Delta _{r-\hat{y}}-\Delta
_{r+\hat{y}})/N$, %
and $d_{xy}$ wave pairing operator $\Delta_{d_{xy}}=\sum_{r}(\Delta
_{r+\hat{x}+\hat{y}}+\Delta _{r-\hat{x}-\hat{y}}-\Delta
_{r+\hat{x}-\hat{y}}-\Delta _{r-\hat{x}+\hat{y}})/N$. %
The respective pair coordinates are shown in Fig. \ref{fig:Pairconf}.

Thus, we have,
\begin{eqnarray}
\lbrack H,\Delta _{s}^{cc}] &=&t_{\perp}\Delta _{s\ast
}^{cc}+t_{\parallel}\Delta _{d_{x^{2}-y^{2}}}^{cc}+t_{3}\Delta
_{s_{xy}}^{cc}-t_{4}\Delta _{d_{xy}}^{cd}  \nonumber \\
&-&(U-2\mu +2U_{cd}n_{r,d})\Delta _{s}^{cc} ~,
\label{CommunicationEquality1} \\
 \lbrack H,\Delta _{s}^{dd}]
&=& t_{\perp}\Delta _{s\ast }^{dd}-t_{\parallel}\Delta
_{d_{x^{2}-y^{2}}}^{dd}+t_{3}\Delta
_{s_{xy}}^{dd}-t_{4}\Delta _{d_{xy}}^{dc}  \nonumber \\
&-&(U-2\mu +2U_{cd}n_{r,c})\Delta _{s}^{dd} ~. \label{CommunicationEquality2}
\end{eqnarray}%
Since there are inter-orbital hopping in the original Hamiltonian (\ref%
{MainHamiltonian}), inter-orbital pairings appear on the right hand side of (%
\ref{CommunicationEquality1}) and (\ref{CommunicationEquality2}). Furthermore,
for quartic terms such as $n_{r,d}\Delta_{r}^{cc}$, we treat them in the mean
field sense by taking number operators $n_{r}$ as its average $\langle
n_{r}\rangle$, assuming that there is no charge-density wave (CDW). Even though
CDW occurs, the strong Coulomb repulsion will destroy on site s wave pairing,
i.e., $\Delta _{s}$=0, and it will not affect our conclusion.
Thus, in the followings, we treat the number operators for two orbitals as $%
\langle n_{r,d}\rangle =\langle n_{r,c}\rangle \equiv \langle n\rangle $ in
all quartic terms. Make a linear combination of (\ref{CommunicationEquality1}%
) and (\ref{CommunicationEquality2}),
\begin{eqnarray}
\big[H,\Delta _{s}^{cc}-\Delta _{s}^{dd}\big]&=&t_{\perp}(\Delta
_{s\ast }^{cc}-\Delta _{s\ast }^{dd}) +t_{\parallel}(\Delta
_{d_{x^{2}-y^{2}}}^{cc}+\Delta _{d_{x^{2}-y^{2}}}^{dd}) \nonumber \\
+ t_{3}(\Delta _{s_{xy}}^{cc}-\Delta _{s_{xy}}^{dd})&-&(U-2\mu +2U_{cd}\langle
n\rangle )(\Delta _{s}^{cc}-\Delta _{s}^{dd}) ~.
\end{eqnarray}%
To express the form more explicitly, with the help of Nambu representation
$\phi _{\sigma }(k)=(c_{-k,\sigma },d_{-k,\sigma })$, the pairing gap in
momentum and orbital space can be expressed as $\sum_{k\alpha \beta }\phi
_{\alpha ,\sigma }(-k)f(k)(\tau _{i})_{\alpha \beta }\phi _{\beta
,\bar{\sigma}}(k)$, where $f(k)$ is symmetry factor in momentum space,
\begin{eqnarray}
\lbrack H,\tau _{3}] &\sim &t_{\perp}(\cos k_{x}+\cos k_{y})\tau
_{3} +t_{\parallel}(\cos k_{x}-\cos k_{y})\tau _{0} \nonumber \\
&+&t_{3}(\cos k_{x}\cos k_{y})\tau _{3}-(U-2\mu +2U_{cd}\langle
n\rangle )\tau _{3} \label{CommunicationEquality4}
\end{eqnarray}%
There are four types of pairing patterns on the right side of equalities (\ref%
{CommunicationEquality4}), listed as No. 1-4 in Table \ref{intra-orbitaltable}.

\begin{table}[t]
\caption{Potential pairing basis matrices under different irreducible
representations of the model, which are classified into three groups in the
table. The first column is the index number, the second and the third columns
denotes the representations and the basis matrix functions respectively. The
parities of spins [singlet(S) or triplet(T)] and orbitals [symmetric(s) or
antisymmetric(a)] are shown in the forth and the last
columns respectively.}%
\begin{ruledtabular}
\begin{tabular}{c c c c c c c c }
& No. & IR & Basis & Spin & Orbital \\
\hline
& 1 & $B_{1}$ & $(\cos k_x +\cos k_y )\tau_3 $ & S & s \\
& 2 & $B_{1}$ & $(\cos k_x -\cos k_y )\tau_0 $ & S & s\\
& 3 & $B_{1}$ & $(\cos k_x \cos k_y )\tau_3$ & S & s \\
& 4& $B_{1}$ & $\tau_3$ & S & s  \\
\hline
& 5 & $B_{2}$ & $(\cos k_x+\cos k_y )\tau_1$ & S & s  \\
& 6 & $B_{2}$ & $(\cos k_x \cos k_y )\tau_1$ & S & s  \\
& 7 & $B_{2}$ & $(\sin k_x \sin k_y )\tau_0$ & S & s \\
& 8& $B_{2}$ & $ \tau_1$ & S & s  \\
\hline
& 9 & $A_{2}$ & $(\cos k_x +\cos k_y )i\tau_2 $ & T & a  \\
& 10 & $A_{2}$ & $(\cos k_x \cos k_y )i\tau_2$ & T & a \\
& 11 & $A_{2}$ & $ i\tau_2  $ &T  & a
 \label{intra-orbitaltable}
\end{tabular}
\end{ruledtabular}
\end{table}

\emph{Inter-orbital pairing}---When the inter-orbital pairing is taken into
consideration, more pairings are involved. Define on site inter-orbital pairing
as $\Delta _{r}^{cd}=c_{r,\uparrow }d_{r,\downarrow }$, %
$\Delta_{r}^{dc}=d_{r,\uparrow }c_{r,\downarrow }$, and NN or NNN inter-orbital
pairing operators in a similar manner,
\begin{eqnarray}
&&\big[H,\Delta _{s}^{cd}+\Delta _{s}^{dc}\big]=t_{\perp}(\Delta
_{s\ast }^{cd}+\Delta _{s\ast }^{dc})+t_{3}(\Delta
_{s_{xy}}^{cd}+\Delta _{s_{xy}}^{dc})-t_{4}(\Delta
_{d_{xy}}^{cc}+\Delta _{d_{xy}}^{dd}) \nonumber \\
&+&\bigg [\big(2\mu -U-2U_{cd}+\frac{3J_{H}}{4}\big)\langle n\rangle -\frac{%
3J_{H}}{4}\bigg](\Delta _{s}^{cd}+\Delta _{s}^{dc}) ~.
\end{eqnarray}%
The pairing gap in momentum and orbital space,
\begin{eqnarray}
\lbrack H,\tau _{1}] &\sim &t_{\perp}(\cos k_{x}+\cos k_{y})\tau
_{1}  \nonumber \\
&+&t_{3}(\cos k_{x}\cos k_{y})\tau _{1}-t_{4}(\sin k_{x}\sin k_{y})\tau _{0}
\nonumber \\
&+&\bigg [\big(2\mu -U-2U_{cd}+\frac{3J_{H}}{4}\big)\langle n\rangle -\frac{%
3J_{H}}{4}\bigg]\tau _{1} ~. \label{CommunicationEquality5}
\end{eqnarray}%
Similarly, with the definition of spin triplet pairings
$\bar{\Delta}_{\vec{\delta} }^{cd}=c_{r\uparrow }d_{r+\vec{\delta}\downarrow
}+c_{r\downarrow }d_{r+\vec{ \delta}\uparrow }$,
$\bar{\Delta}_{\vec{\delta}}^{dc}=d_{r\uparrow }c_{r+\vec{\delta}\downarrow
}+d_{r\downarrow }c_{r+\vec{\delta}\uparrow }$, we have

\begin{eqnarray}
\big[H,\Delta _{s}^{cd}-\Delta _{s}^{dc}\big] &=& t_{\perp}(\bar{\Delta}_{s\ast
}^{cd}-\bar{\Delta}_{s\ast }^{dc}) +t_{3}(\bar{\Delta
}_{s_{xy}}^{cd}-\bar{\Delta}_{s_{xy}}^{dc})  \nonumber \\
+\bigg[\big(2\mu -U&-&2U_{cd}-\frac{J_{H}}{4}\big)\langle n\rangle +\frac{%
J_{H}}{4}\bigg](\Delta _{s}^{cd}-\Delta _{s}^{dc})
\end{eqnarray}%
\begin{eqnarray}
\lbrack H,i\tau _{2}] &\sim &t_{\perp}(\cos k_{x}+\cos
k_{y})i\tau _{2}+t_{3}(\cos k_{x}\cos k_{y})i\tau _{2}  \nonumber \\
&+&\bigg[\big(2\mu -U-2U_{cd}-\frac{J_{H}}{4}\big)\langle n\rangle +\frac{%
J_{H}}{4}\bigg]i\tau _{2}  \label{CommunicationEquality6}
\end{eqnarray}%
There are four types of pairing on the right side of equalities (\ref%
{CommunicationEquality5}) listed as No. 5-8 and three in (\ref%
{CommunicationEquality6}) listed as No. 9-11 in Table \ref{intra-orbitaltable},
where the gap behaviors are also enumerated.

\emph{Inequality and analysis}---With the use of the above commutation
relations, let us establish the sufficient condition for the coexistence of two
long-range orders rigorously. We should take advantage of off-diagonal long
range theory \cite{Yang-RMP-1962} and generalizing the approach of Ref. \cite%
{GSTian-JPA-1993} to obtain a strict proof.

For a general operator $G_{k}$ defined on lattice $\Lambda$, it has a
long-range order if and only if (iff) its reduced-density matrix in $k$ space
satisfies
\[
\langle \Psi _{0}(\Lambda )|G_{k}^{\dagger }G_{k}|\Psi _{0}(\Lambda )\rangle
\geq \lambda N_{\Lambda },\quad \lambda >0 ~,
\]%
at a certain $k=k_{0}$ point as $N_{\Lambda }\rightarrow \infty $ with fixed
density. Assuming an operator $A_{k}$ satisfies the following commutation
relation
\begin{equation}
\lbrack H_{\Lambda },A_{k}] = \alpha B_{k} + \beta C_{k} + \gamma D_{k}\equiv
Q_{k} ~, \label{eq:operatorcomm}
\end{equation}%
then we can prove that
\begin{eqnarray}
&&\langle \Psi _{0}(\Lambda )|Q_{k}^{\dagger }Q_{k}|\Psi _{0}(\Lambda
)\rangle \leq m(Q_{k})m(A_{k})  \label{inequalityofcorrelation2} ~, \\
&&m(Q_{k})=\sqrt{\langle \Psi _{0}(\Lambda )|[{{{{Q_{k}^{\dagger
},[H_{\Lambda },Q}}}}_{k}{]}]|\Psi _{0}(\Lambda )\rangle } ~, \\
&&m(A_{k})=\sqrt{\langle \Psi _{0}(\Lambda )|[{{{{A_{k}^{\dagger },[H_{\Lambda
},A}}}}_{k}{]}]|\Psi _{0}(\Lambda )\rangle } ~.
\end{eqnarray}%
Here $m(Q_{k})$ and $m(A_{k})$ are quantities of order $O(1)$ as $N_{\Lambda
}$ tends to infinity. Therefore, the correlation function of $Q_{k}$ is at
most a quantity of order $O(1)$. Then expanding the inequality (\ref%
{inequalityofcorrelation2}) and applying the Cauchy-Schwarz inequality, we have
\begin{eqnarray}
&&\sum_{\mu }|\lambda _{\mu }|^{2}\langle \Psi _{0}(\Lambda
)|G_{k}^{\mu \dagger }G_{k}^{\mu }|\Psi _{0}(\Lambda )\rangle
\leq m(Q_{k})m(A_{k})  \nonumber \\
&+&\sum_{\mu \nu }2|\lambda _{\mu }||\lambda _{\nu }|%
\sqrt{\langle \Psi _{0}(\Lambda )|G_{k}^{\mu \dagger }G_{k}^{\mu }|\Psi
_{0}(\Lambda )\rangle \langle \Psi _{0}(\Lambda )|G_{k}^{\nu \dagger
}G_{k}^{\nu }|\Psi _{0}(\Lambda )\rangle} \nonumber
\label{inequalityofcorrelation1}
\end{eqnarray}%
where $G_{k}^{\mu }$ represents $B_{k}(C_{k},D_{k})$ in Eq. (\ref%
{eq:operatorcomm}). Now let us assume that $\Psi _{0}(\Lambda )$ has a kind
of long-range order, say $B_{k}$, then
\begin{equation}
\langle \Psi _{0}(\Lambda )|B_{k}^{\dagger }B_{k}|\Psi _{0}(\Lambda )\rangle
\geq \lambda _{B}N_{\Lambda },\quad \lambda _{B}>0  ~, \label{eq:rigresultb}
\end{equation}%
as $N_{\Lambda }\rightarrow \infty $. Since the left hand side of inequality
(\ref{inequalityofcorrelation1}) is a quantity of order $O(N_{\Lambda })$,
and other operator's correlation function is, at most, a quantity of order $%
O(1)$ in the thermodynamic limit, the right hand side of the inequality can
be, at most, a quantity of order $O(\sqrt{N_{\Lambda }})$. Therefore, $\Psi
_{0}(\Lambda )$ must have another long-range order of, say $C_{k}$, that is,
\begin{equation}
\langle \Psi _{0}(\Lambda )|C_{k}^{\dagger }C_{k}|\Psi _{0}(\Lambda )\rangle
\geq \lambda _{C}N_{\Lambda },\quad \lambda _{C}>0.  \label{eq:rigresultc}
\end{equation}%
The conclusion is that either all orders are absent or at least two
long-range orders must be present simultaneously in the ground state $\Psi
_{0}(\Lambda )$ of the Hamiltonian $H_{\Lambda }$.

Applying the above rigorous result to Eq. (\ref{MainHamiltonian}), we have a
basic conclusion that orders of the same group listed in Table
\ref{intra-orbitaltable}, say No.1-4, should either are all absent, or at least
two of them coexist. Specifically, we can refine this conclusion as follows.

Firstly, the strong on-site Coulomb interaction and large Hund's coupling
suppresses all the states involving $s$ wave pairing, then No. 4 (symmetric
intra-orbital $s$), 8 (symmetric inter-orbital $s$), 11 (anti-symmetric
inter-orbital $s$) will be unfavorable.

Secondly, Eq. (\ref{CommunicationEquality4}) shows that all order parameters in
the first group of Table \ref{intra-orbitaltable} are of spin singlet,
intra-orbital, and even parity pairing symmetry. Nevertheless, our rigorous
results [Eqs. (\ref{eq:rigresultb} and \ref{eq:rigresultc})] require that
orders of No 1-3 should either vanish simultaneously or at least two of them
coexist. If two of them coexist, the magnitudes of the two order parameters are
determined by the multiplying their pairing functions ($f_{1}(k)$, $
f_{2}(k)$). So if $s^{\ast }$ and $s_{xy}$ coexist, $f_{1}(k)=(\cos k_{x}+\cos
k_{y})$, $f_{3}(k)=\cos k_{x}\cos k_{y}$. Their overlap becomes dominated
around the hole pocket about $\Gamma$ points in the Brillouin zone, but is very
small around the electron pocket about $M$ points. While if $d_{x^{2}-y^{2}}$
and $s_{xy}$ coexist, $f_{2}(k)=(\cos k_{x}-\cos k_{y})$, $f_{3}(k)=\cos
k_{x}\cos k_{y}$, their product has an enhanced contribution from the electron
pocket about $M$ points, but is suppressed from hole pocket about $\Gamma$
points. However, the coexistence of $s^{\ast }$ and $d_{x^{2}-y^{2}}$ is not
favorable because of their overlap is very tiny around both $M$ and $\Gamma $
points.

Thirdly, in the second group of Table \ref{intra-orbitaltable}, there are
No. 5 (symmetric inter-orbital $s^{\ast }$), 6 (symmetric inter-orbital $%
s_{xy}$), and 7 (intra-orbital $d_{xy}$) left on the right side of (\ref%
{CommunicationEquality5}). Since $f_{7}(k)=\sin k_{x}\sin k_{y}$ is peaked
around ($\pi /2$,$\pi /2$) , if No. 7 coexists with No. 5 or No. 6, $%
f_{7}(k) $ has a tiny contribution at $M$ and $\Gamma $ points and does
support the current fermi surface topology, therefore No. 7 is not favored.
So the remaining $s^{\ast }$ and $s_{xy}$ of spin singlet must be either
absent together or coexist simultaneously. According to the behavior of
symmetry factors $f_{5(6)}(k)$, the coexistence of $s^{\ast }$ and $s_{xy}$
is competitive only around hole pocket about $\Gamma $ points, but not
favored in electrons pocket about $M$ points.

Fourthly, regarding with the third group including No. 9 (antisymmetric
inter-orbital $s^{\ast }$) and No. 10 (antisymmetric inter-orbital $s_{xy}$) on
the right hand side of (\ref{CommunicationEquality6}), our rigorous results
[Eqs. (\ref{eq:rigresultb}) and (\ref{eq:rigresultc})] impose that either both
of them are absent or coexist. Because both of them carry antisymmetric orbital
parity and become gapless in excitation spectrum, which are
inconsistent with the experimental evidence of nodal gap \cite%
{Matano-arXiv08060249,HHWen}, the chance of coexistence seems slim.

Finally, in the weak coupling limit, two orbitals' energy splitting might lead
to a mismatch of inter-orbital pairing in momentum space with opposite sign,
instead of pairing between two different $|k|$s. That is the piling up of
low-energy density of state in the gapless SC state will lead to a
Fulde-Ferrel-Larkin-Ovchinnikov state with magnetic ordering, and does not
create SC instability. In this sense, the orbital antisymmetric pairing state
such as Nos. 9 and 10 might be ruled out. Moreover, other inter-band pairings,
such as Nos. 5 and 6 in Table \ref{intra-orbitaltable} are also not favorable
according to the analysis based on the FFLO state. Then we arrive at our
further conclusion that around half filling, in the electron doping region, the
system will favor coexistence of $d_{x^{2}-y^{2}}$ and $s_{xy}$ waves pairing,
while in the hole doping region, the system might prefer to have $s^{\ast }$
and $s_{xy}$ waves pairing.

To summarize, we have built some identities based on a two-orbital model, and
obtained constraints on a few possible pairings. Our results provide more
information than group theory classification. According to the sufficient
condition for coexistence of two superconducting orders and resorting to
physical consideration, we propose the most favorable pairings around half
filling. Although our discussions are based on a two-orbital model, it is
straightforward to generalize the strategy to other Hamiltonians even if more
orbitals are involved. In principle, we have not ruled out the spatial odd
parity pairing, e.g., $p$ wave, which can be achieved by commutation between an
odd parity pairing operator and Hamiltonian, and the sufficient condition of
coexistence of the odd parity pairings is still applicable. Nevertheless, they
do not get along well with our fermi surface topology analysis given above.

We are grateful to Dr. Yi Zhou for valuable comments. This works is supported
HKRGC (Project Nos. CUHK 402205 and HKU3/05C).

\end{document}